\documentclass[aps,prl,twocolumn,numerical,superscriptaddress,nofootinbib,showpacs,showkeys,hyperref,longbibliography]{revtex4-1}

\usepackage{amsmath,amssymb,amsfonts}
\usepackage[colorlinks=true,plainpages=true]{hyperref}

\usepackage[dvipdfm]{graphicx}
\usepackage{bmpsize}

\newcommand{\pT}{\mathrm{p_{\mathrm{T}}}}

\begin{document}
\title{
Tracing the Evolution of Nuclear Excitation at the Electron-Ion Collider
}

\medskip

\author{Niseem~Magdy}
\email{niseem.abdelrahman@tsu.edu}
\affiliation{Department of Physics, Texas Southern University, Houston, TX 77004, USA}
\affiliation{Physics Department, Brookhaven National Laboratory, Upton, New York 11973, USA}


\begin{abstract}
We investigate the evolution of nuclear excitation in electron–nucleus ($e+A$) collisions at the upcoming Electron-Ion Collider (EIC) using the BeAGLE event generator. Leveraging the EIC's unique collider kinematics, we demonstrate the remarkable capability to separate distinct nuclear reaction stages, hard scattering, intranuclear cascade, and nuclear de-excitation, in the laboratory frame. Our systematic analysis reveals that event-by-event fluctuations in the mean transverse momentum ($\kappa$) are highly sensitive to the intranuclear cascade formation time and nuclear geometry, while minimally affected by variations in electron beam energy. These findings establish $\kappa$ as a robust observable for constraining nuclear excitation mechanisms, providing critical benchmarks for future EIC experiments, and guiding theoretical advancements in nuclear transport modeling.
\end{abstract}

\maketitle

Understanding the evolution of nuclear excitation, specifically, how energy is deposited, redistributed, and ultimately dissipated following high-energy lepton-ion interactions, remains an active area of research~\cite{Bertulani:2003, Bertsch:1988ik, Buss:2011mx, Jucha:2024ltd, Larionov:2018igy}. Despite decades of investigation, several aspects of these processes are still not fully understood. Nuclear excitation mechanisms govern a wide range of phenomena, including particle emission, fission, fragmentation, and $\gamma$-ray de-excitation~\cite{Schmidt:2018hwz, Schunck:2015hxw, Buss:2011mx, Charity:2010, Bondorf:1995ua}. A central challenge is the clear separation of the various dynamical stages that contribute to nuclear excitation and subsequent decay~\cite{Boudard:2002yn}.

 \begin{figure}[!h]
 \vskip -0.26cm
 \centering{
 \includegraphics[width=1.0\linewidth,angle=0]{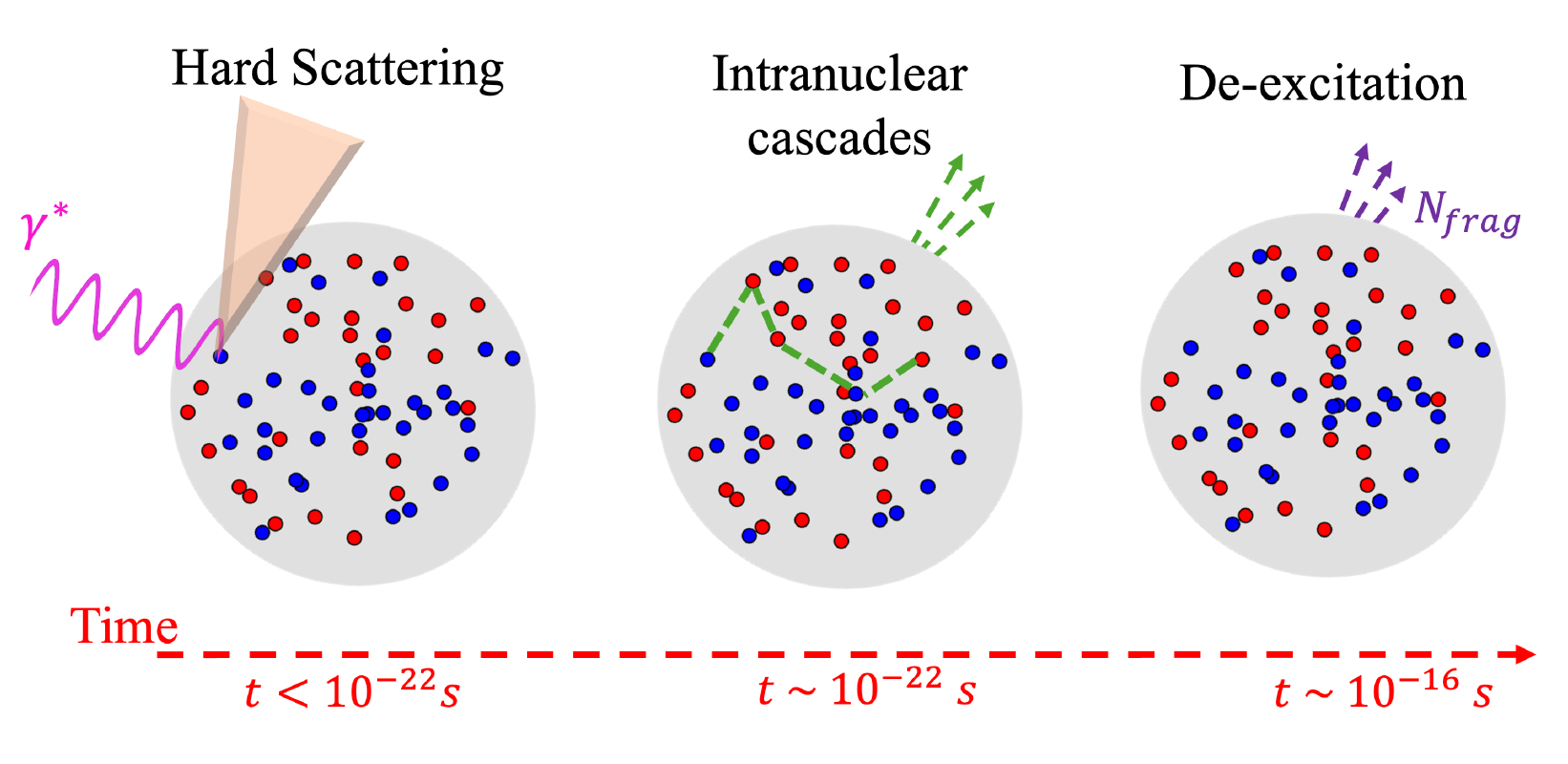}
\vskip -0.36cm
 \caption{
Cartoon illustrating the reaction stages expected in lepton/photon nucleus collisions.
 \label{fig:0}
 }
 }
 \end{figure}

\begin{figure}[!h]
 \centering{
 \includegraphics[width=1.00\linewidth,angle=0]{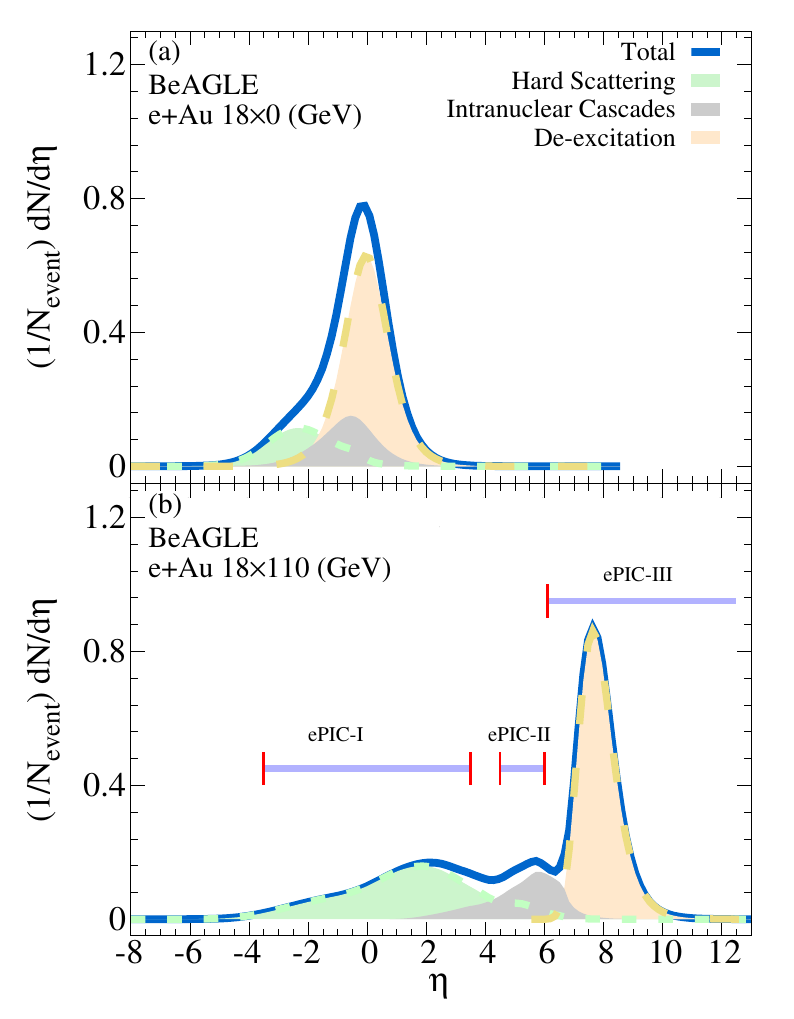}
 \vskip -0.36cm
 \caption{
The rapidity distribution of particle multiplicity ($dN/d\eta$) of the $e$+$Au$ collisions at 18$\times$0 (GeV), illustrates a fixed-target configuration, panel (a) and 18$\times$110 (GeV) panel (b) from the BeAGLE model. The shaded areas represent the different contributions, and the horizontal lines represent the $\eta$ acceptance given by the ePIC experiment at the EIC~\cite{AbdulKhalek:2021gbh, ePIC}.
 \label{fig:1}
 }
 }
 \end{figure}

In lepton- or photon–nucleus collisions, the reaction typically proceeds through three distinct stages in the nuclear rest frame, as illustrated in Fig.~\ref{fig:0}: (i) an instantaneous hard scattering~\cite{Piller:1995kh, Arneodo:1994wf}, (ii) a rapid intranuclear cascade occurring within approximately $10^{-22}$~s~\cite{Cugnon:1982qw}, and (iii) a longer-timescale de-excitation process lasting up to $10^{-16}$~s~\cite{Weisskopf:1937zz, Charity:2010, Bondorf:1995ua}. The duration and characteristics of each stage are expected to depend on both the properties of the target nucleus and the specific collision kinematics~\cite{Mathews:1982zz, Magdy:2024thf}. Advancing our understanding of nuclear excitation, therefore, depends critically on the ability to distinguish between the different reaction stages and to clarify how each stage influences the others. In particular, the interplay between the intranuclear cascade phase and the thermalized compound nucleus de-excitation remains a key open question in the field~\cite{Boudard2002, Lehr2000}. Addressing this issue has been challenging, in large part due to limitations in experimental methods; most studies to date have been restricted to fixed-target configurations~\cite{ALADIN:2002, HADES:2011, Cerizza:2016}. These limitations are exemplified in Fig.~\ref{fig:1}(a), which shows the rapidity dependence of particle multiplicity ($dN/d\eta$) for e+Au collisions at $10\times0$ GeV, with substantial overlap among the hard scattering, intranuclear cascade, and nuclear de-excitation processes that complicates efforts to clearly distinguish their contributions.

Beyond the constraints of fixed-target experiments, a significant challenge lies in the absence of robust experimental constraints on how the intranuclear cascade process modulates the breakup of excited nuclei. This modulation generally depends on the intranuclear cascade formation time ($\tau_f$), defined as the interval between the production of a secondary hadron and the onset of its ability to interact further within the nuclear medium~\cite{Lehr2000, Bass:1998ca}. For very short $\tau_f$, secondary hadrons interact almost immediately after their production, resulting in a dense intranuclear cascade and high excitation energies; for very long $\tau_f$ (i.e., $\tau_f \gtrsim R_A/c$), hadrons can escape the nucleus before re-interacting, thereby suppressing the cascade and resulting in minimal nuclear excitation. Meeting these challenges requires (i) identifying experimental-kinematic regions that can isolate different reaction stages, and (ii) establishing observables that can constrain $\tau_f$~\cite{Larionov:2018igy}.

The upcoming Electron-Ion Collider (EIC)~\cite{AbdulKhalek2021, Accardi:2012qut} presents a unique opportunity to investigate the evolution of nuclear excitation with collider kinematics. In collider-based $e+A$ interactions, the three reaction stages illustrated in Fig.~\ref{fig:0} are expected to be well separated in the laboratory frame, as shown in Fig.~\ref{fig:1}. The illustration given in Fig.~\ref{fig:1} (b) demonstrates that it is possible to isolate kinematic regions dominated by each reaction stage~\cite{Chang:2022, Magdy:2024thf}. In Fig.~\ref{fig:1} panel (b), the regions labeled ePIC-I, ePIC-II, and ePIC-III correspond to the acceptance ranges of the three primary subsystems anticipated for the $ePIC$ detector~\cite{ePIC}, which are the focus of this study.
Beyond these intrinsic capabilities of the EIC, we propose utilizing event-by-event fluctuations in the mean transverse momentum ($M(\pT)$) as a measurement to constrain the $\tau_f$ effect in $e+A$ collisions~\cite{Adams:2005ka, ALICE:2019nqi, Zhang:2025yyd}. Specifically, $M(\pT)$ fluctuations encode information about the origin of emitted particles: the de-excitation stage tends to yield more thermal-like, isotropic, and statistically narrow momentum distributions, whereas the intranuclear cascade phase leads to broader and more irregular distributions due to the stochastic and non-thermal character of secondary particle interactions. Thus, $M(\pT)$ fluctuations are expected to serve as a sensitive probe of the underlying particle production mechanism, which is anticipated to depend on $\tau_f$. Consequently, conducting such studies at the EIC will be essential for refining nuclear excitation models and for achieving deeper insight into the structure and dynamics of the target nucleus~\cite{Mathews:1982zz, Magdy:2024thf, AbdulKhalek2021, Accardi:2012qut}.


This study employs version 1.03 of the Benchmark eA Generator for LEptoproduction (\textsc{BeAGLE})~\cite{Chang:2022hkt} Monte Carlo simulation code. \textsc{BeAGLE} is a hybrid event generator that integrates several well-established codes, including \textsc{DPMJet}~\cite{Roesler:2000he}, \textsc{PYTHIA6}~\cite{Sjostrand:2006za}, \textsc{PyQM}~\cite{Dupre:2011afa}, \textsc{FLUKA}~\cite{Bohlen:2014buj, Ferrari:2005zk, Battistoni:2015epi}, and \textsc{LHAPDF5}~\cite{Whalley:2005nh}, to comprehensively describe high-energy lepton-nucleus scattering processes. In this framework, \textsc{DPMJet} governs hadron production and their interactions with the nucleus via intranuclear cascades~\cite{Roesler:2000he}, while \textsc{PYTHIA6} simulates the underlying partonic interactions and subsequent fragmentation. The \textsc{PyQM} model provides the geometric nucleon density distribution and implements the Salgado-Wiedemann quenching weights to describe partonic energy loss in the medium~\cite{SW:2003}. \textsc{FLUKA} simulates the decay of the excited nuclear remnant, encompassing nucleon and light ion evaporation, nuclear fission, Fermi breakup, and photon emission de-excitation~\cite{Bohlen:2014buj, Ferrari:2005zk, Battistoni:2015epi}. Additionally, \textsc{BeAGLE} incorporates features such as multi-nucleon scattering (shadowing) and improved modeling of the Fermi momentum distribution of nucleons in nuclei~\cite{Chang:2022hkt}.

The present analysis specifically examines the interplay between intranuclear cascades, modeled by \textsc{DPMJet}, and nuclear thermal de-excitation processes, simulated by \textsc{FLUKA}, achieved by varying the formation time parameter $\tau_{f}$.  In \textsc{DPMJet}, intranuclear cascades represent a secondary stage of interactions following the primary collision, where secondary particles produced in the initial event traverse the nuclear medium and undergo successive scatterings with bound nucleons~\cite{Roesler:2000he}. These cascades set the initial conditions for the subsequent evolution of the compound nucleus, which then de-excites according to the mechanisms described in \textsc{FLUKA}~\cite{Battistoni:2015epi, Bohlen:2014buj}.

Using the BeAGLE model, we generated 50 million events for each presented scenario to investigate the event-by-event transverse momentum fluctuations. The proposed observable in this study is the variance of the event-by-event transverse momentum distribution, denoted by $\kappa$~\cite{CERES:2003sap, Gavin:2003cb, Bhatta:2021qfk}, which quantifies the strength of dynamical fluctuations relative to the mean transverse momentum, $M(p_T)$, for a selected class of particles. The quantity $\kappa$ is formally defined as follows:
\begin{align}
\kappa = \frac{C}{M(\pT)^{2}},
\end{align}
where $C$ is the second-order momentum correlation. Both, $C$ and $M(\pT)$ are given as; 
{\small 
\begin{align}
C = \frac{1}{\sum^{n_{e}}_{k=1} N^{pairs}_{k}}  \sum^{n_{e}}_{k=1} \sum^{n_{p}}_{i=1} \sum^{n_{p}}_{j=i+1} ( \pT_i - M(\pT) ) ( \pT_j - M(\pT) ),
\end{align}
\begin{align}
M(\pT) = \frac{1}{\sum^{n_{e}}_{k=1} N_{k}}  \sum^{n_{e}}_{k=1} \sum^{n_{p}}_{i=1}  \pT_i.
\end{align}
}
where $n_e$ is the total number of events and $n_p$ is the number of particles in such an event.

 \begin{figure}[!h]
 \centering{
 \includegraphics[width=1.0\linewidth,angle=0]{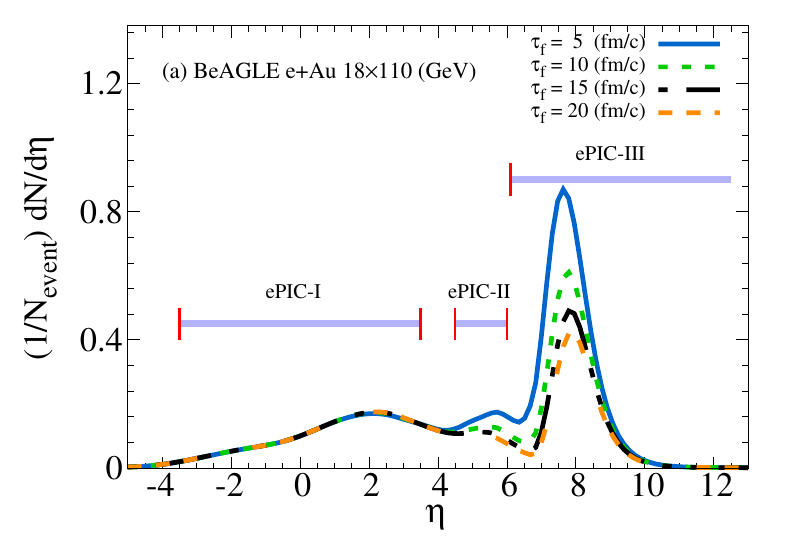}
 \vskip -0.36cm
 \caption{
Panel (a) shows the $dN/d\eta$ distribution of the $e$+$Au$ collisions at 18$\times$110 (GeV) from the BeAGLE model at different formation times.
 \label{fig:2}
 }
 }
 \end{figure}
 
Before discussing our primary observable $\kappa$, we found it instructive to first examine the dependence of the particle multiplicity distribution, \(dN/d\eta\), on the formation time~\cite{Zheng:2014cha}. Figure~\ref{fig:2} presents the \(dN/d\eta\) distribution for \(e+\mathrm{Au}\) collisions at 18$\times$110~GeV, as simulated with the \textsc{BeAGLE} model for different formation times. The results indicate no significant sensitivity to changes in formation time at mid-rapidity (ePIC-I range), where hard scattering processes dominate. In contrast, a clear increase in \(dN/d\eta\) is observed with decreasing formation time at forward rapidity (ePIC-II range) and far-forward rapidity (ePIC-III range), where the intranuclear cascade and nuclear de-excitation processes, respectively, are dominant. However, because both ePIC-II and ePIC-III regions show increased particle production with decreasing formation time, it remains challenging to disentangle the effects arising from intranuclear cascades and those from de-excitation. This underscores the need for an additional observable capable of distinguishing between these two contributions.

 \begin{figure}[!h]
 \centering{
 \includegraphics[width=1.0\linewidth,angle=0]{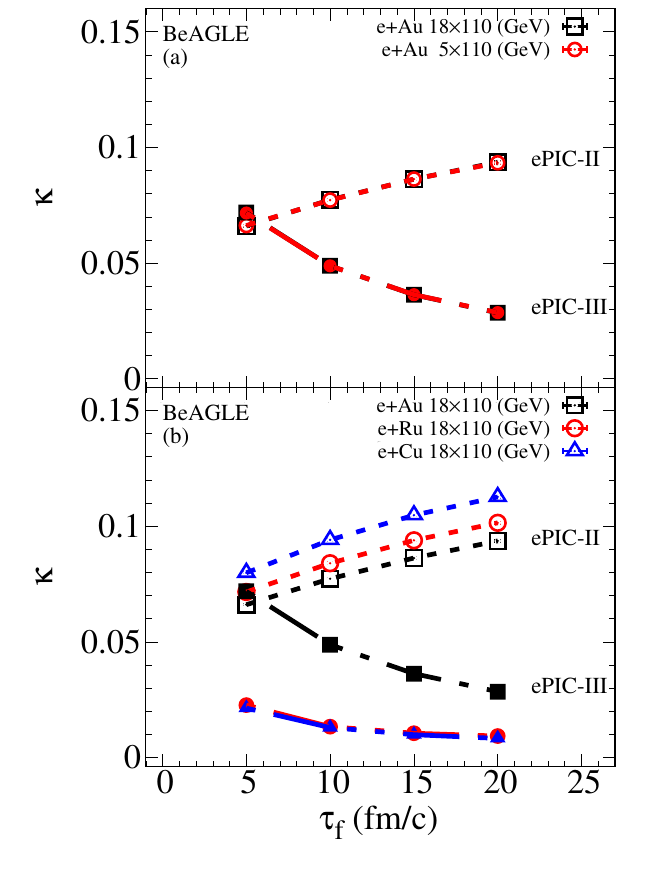}
\vskip -0.40cm
 \caption{
 The energy and system size dependence of \(\kappa\) as a function of formation time from the BeAGLE model for ePIC-II and ePIC-III regions. The lines are drawn to guide the eye.
 \label{fig:3}
 }
 }
 \end{figure}

Figure~\ref{fig:3} displays \(\kappa\) as a function of formation time for various configurations. Panel (a) shows results for \(e+\mathrm{Au}\) collisions at 18$\times$110~GeV and 5$\times$110~GeV, while panel (b) presents results for \(e+\mathrm{Au}\), \(e+\mathrm{Ru}\), and \(e+\mathrm{Cu}\) at 18$\times$110~GeV. The results are shown for the ePIC-II and ePIC-III rapidity selections, where the intranuclear cascade and nuclear de-excitation processes dominate, respectively.
Our findings reveal that \(\kappa\) increases with $\tau_f$ in the intranuclear cascade-dominated region and decreases with $\tau_f$ in the de-excitation-dominated region. This contrasting behavior can be understood in terms of the distinct dynamical responses of each process to the formation time. For short $\tau_f$, secondary hadrons produced by the intranuclear cascade interact more frequently with the nuclear medium, resulting in more cascades and higher excitation energy. Conversely, for longer $\tau_f$, secondary hadrons experience minimal interaction with the nuclear medium, leading to reduced excitation energy and consequently, lower de-excitation.

In Fig.~\ref{fig:3}, panel (a), our results indicate that for a fixed nuclear target, \(\kappa\) exhibits minimal dependence on the electron beam energy across both ePIC-II and ePIC-III rapidity regions. This energy independence suggests that the particle production dynamics encoded in the intranuclear cascade and nuclear de-excitation models within \textsc{BeAGLE} are relatively insensitive to the initial electron energy over the examined range. Specifically, varying the electron energy from 5~GeV to 18~GeV has minimal impact on \(\kappa\), indicating that the dominant source of fluctuations is governed by nuclear geometry and medium interactions rather than by the electron energy itself.
In contrast, Fig.~\ref{fig:3}, panel (b), demonstrates a pronounced sensitivity of \(\kappa\) to the size of the target nucleus. As the system size decreases from Au to Ru to Cu, the magnitude and formation-time dependence of \(\kappa\) change significantly. This highlights the importance of geometric and nuclear density effects on the evolution of momentum fluctuations. Larger nuclei provide a more extended medium for secondary particle interactions, which can enhance or suppress fluctuations (ePIC-II or ePIC-III regions) depending on the process type and formation time. Thus, system size emerges as a critical factor in shaping the observed \(\kappa\) trends and offers an essential means for disentangling the contributions of intranuclear cascade and nuclear de-excitation mechanisms in future studies.

 \begin{figure}[t]
 \centering{
 \includegraphics[width=1.0\linewidth,angle=0]{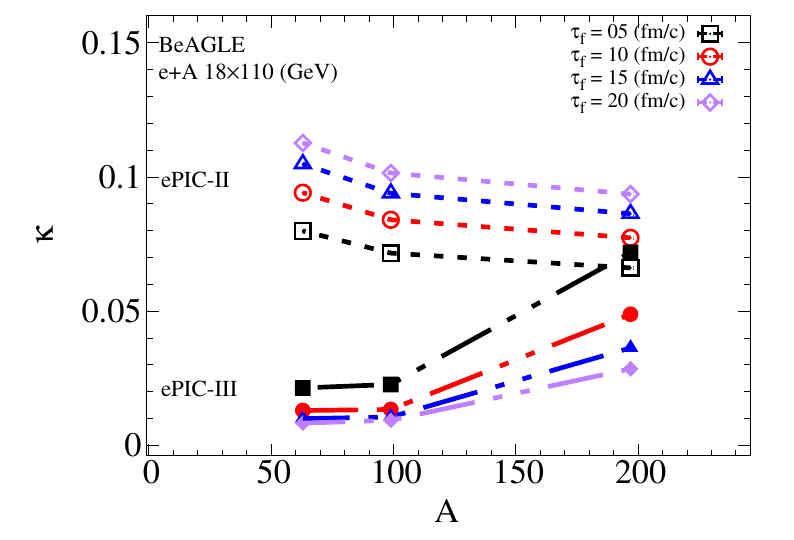}
\vskip -0.40cm
 \caption{
  The system size of the \(\kappa\) as a function of the atomic mass number for several formation times from the BeAGLE model for ePIC-II and ePIC-III regions. The lines are drawn to guide the eye.
 \label{fig:4}
 }
 }
 \end{figure}
A direct measurement of the formation time remains inaccessible in experimental settings. To address this limitation, we present our calculations as a function of the atomic mass number \(A\), which reflects the size and density profile of the target nucleus. This approach enables us to indirectly explore the effects of formation time by examining systematic trends in observables across nuclei of varying sizes.
Figure~\ref{fig:4} shows the behavior of \(\kappa\) for both the intranuclear cascade and de-excitation regions as a function of atomic mass number. We find that \(\kappa\) increases with decreasing atomic mass number for the ePIC-II selection (i.e., the intranuclear cascade-dominated region). In contrast, for the ePIC-III selection (i.e., the de-excitation-dominated region), \(\kappa\) decreases with A, with weaker system-size dependence for small nuclei. In addition, a clear formation time dependence was observed in both regions.  Collectively, these results provide valuable benchmarks for future experimental studies and offer important constraints for the implementation of formation time in theoretical nuclear transport models.

In summary, using the \textsc{BeAGLE} event generator, we explored the potential of the unique kinematic environment at the upcoming Electron-Ion Collider to separate distinct nuclear reaction stages in electron--nucleus collisions. This study introduces a novel method to constrain the intranuclear cascade formation time, and, consequently, the interplay between the intranuclear cascade and nuclear de-excitation processes, by measuring event-by-event fluctuations in the mean transverse momentum, $\kappa$.
Our calculations reveal that $\kappa$ is highly sensitive to the formation time and the size of the target nucleus, while it shows minimal sensitivity to variations in the electron energy within the studied range. These findings indicate that $\kappa$ is a robust observable for probing the dynamics of nuclear reactions in $e$+$A$ collisions. Moreover, the contrasting system-size dependence of $\kappa$ for intranuclear cascade and nuclear de-excitation processes offers a promising path for disentangling the contributions from these distinct stages at the EIC, where electron--nucleus collisions can be investigated with unprecedented precision.

\section*{Acknowledgments}
The author gratefully acknowledges Rongrong Ma, Tanner Mengel, Wenbin Zhao, and Soren P. Sorensen for their valuable discussions and insights. The author also appreciates the important contributions and discussions within the EIC Rare Isotopes group. This work utilized the computational resources provided by the High-Performance Computing Center at Texas Southern University.




\bibliography{ref}

\end{document}